\documentclass{ws-procs9x6}

\newcommand{\refeq}[1]{(\ref{#1})}

\begin{document}

\title{RENORMALIZATION AND ASYMPTOTIC STATES IN LORENTZ-VIOLATING QFT}

\author{M.\ CAMBIASO}

\address{ Universidad Andres Bello, Departamento de Ciencias F\'isicas, \\
Facultad de
Ciencias Exactas, 
Av. Rep\'ublica 220, Santiago, Chile.\\
$^*$E-mail: mcambiaso@unab.cl}

\author{R.\ LEHNERT}

\address{Indiana University Center for Spacetime Symmetries, \\
Bloomington, IN 47405, USA.\\
E-mail: ralehner@indiana.edu}

\author{R.\ POTTING}

\address{CENTRA, Departamento de F\'\i sica, Universidade do Algarve, \\
8005-139 Faro, Portugal.\\
E-mail: rpotting@ualg.pt}

\begin{abstract}
Radiative corrections in  quantum field theories  
with small departures 
from Lorentz symmetry alter  structural aspects
of the theory, in particular the definition of 
asymptotic single-particle
states. Specifically, the mass-shell condition, 
the standard renormalization procedure 
as well as the Lehmann--Symanzik--Zimmermann reduction formalism are affected.
\end{abstract}

\bodymatter

In the presence of Lorentz breakdown, 
previous analyses involving the properties 
of freely propagating particles
have been performed under the tacit assumption that
the physics of free particles is determined by
the quadratic pieces of the corresponding Lagrangian.
However, 
by analyzing  radiative corrections 
for a sample Lorentz-violating (LV) Lagrangian 
contained in the SME, 
we claim that the 
previous line of reasoning  fails.


We consider 
the  bare gauge-invariant Lagrangian density for single-flavor QED 
within the minimal SME in the presence of 
the $c^{\mu \nu}$ and $\tilde k^{\mu \nu}$ only. Both
are symmetric and traceless and the latter is defined as  a $\tilde k^{\mu \nu} = (k_F)^{\mu \alpha \nu}_{\phantom{\mu \alpha \nu}\alpha}$. 
One-loop multiplicative renormalizability   
of the model is  assumed\cite{KLP}. 
For convenience we write the gauge-fixed bare Lagrangian  
together with appropriate IR regularization 
terms\cite{Stueckelberg}. %
To understand  the external fermion states we need to
consider radiative corrections to the fermion two-point 
function, wherefrom the wave-function renormalization 
``constant'' is read-off as the residue of the one-particle 
pole. As a zeroth-order system on which 
perturbation theory is set up, we choose the full renormalized
quadratic Lagrangian together with the LV part. 
The remaining non-quadratic contributions to the Lagrangian
being taken as perturbations. 
Thus, the proper two-point function adopts the form 
$\Gamma^{(2)}(p) =  \Gamma^\mu p_\mu  -m- \Sigma(p^\mu)$, where $\Gamma^\mu =\gamma^\mu + c^{\mu \nu}\gamma_\nu$ and 
\begin{eqnarray}
\Sigma(p^\mu) &=& \Sigma_{\rm LI}(  / \!\!\!p)
+\Sigma_{\rm LV}(p^2, c^{p}_\gamma, \tilde{k}^{p}_\gamma)
+\delta\Sigma
(p^\mu,c^{\mu\nu},\tilde{k}^{\mu\nu}),
\label{2pt-function2}
\end{eqnarray}
with  $c^{p}_\gamma \equiv c^{\mu \nu} p_\mu \gamma_\nu$
and similarly for $\tilde k^p_\gamma$.
The  first term on the RHS of Eq.\ \refeq{2pt-function2} denotes 
the usual Lorentz-symmetric contributions. The rest  
come from LV terms. The last, however,  contains terms 
that are not
present in the original Lagrangian. To linear order in LV and given that electromagnetic interactions preserve C, P, and T, 
we can write:
\begin{eqnarray}
\Sigma_{\rm LV}(p^2, c^{p}_\gamma, \tilde{k}^{p}_\gamma) &=&
f_2^c(p^2)\, c^{p}_\gamma +f_2^{\tilde k}(p^2)\,\tilde{k}^{p}_\gamma,\\
\delta\Sigma(p^\mu,c^{p}_p,\tilde{k}^{p}_p)&=&
f_3^c(p^2)\, \frac{c^{p}_p}m
+f_4^c(p^2)\, \frac{\slash \!\!\!p c^{p}_p}{m^2} 
+f_3^{\tilde k}(p^2)\,\frac{\tilde{k}^{p}_p}m
+f_4^{\tilde k}(p^2)\,\frac{\slash \!\!\!p\tilde{k}^{p}_p}{m^2}.
\end{eqnarray}
The  functions 
$f^c_i(p^2)$ and $f^{\tilde k}_i(p^2)$ are
calculable to any order in the 
fine-structure constant $\alpha$. Following
Ref.\ \refcite{KL} 
we extract the propagator pole:%
\footnote{ %
The  
dependence of $\Gamma^{(2)}(p)$ on 
$c^p_\gamma$ and $\tilde k^p_\gamma$ introduce
further subtleties. Details of the one-loop calculation  together with
the necessary Feynman rules  can be
found in Ref.\ \refcite{CLP}.}%
%
\begin{equation}
\Gamma^{(2)}(p)=
\mathcal Z_R^{-1} \,\,\bar P(p)+
\bar P(p)\,\, \Sigma_2\bigl(/\!\!\! p,c^p_\gamma,\tilde k^p_\gamma,(c,k)^p_p\bigr)\,\bar P(p),
\label{LV-Gamma-pole}
\end{equation}
which yields the desired property:
$\Gamma^{(2)}(p)^{-1}=\mathcal Z_R\,\,\bar P(p)^{-1}+\mbox{finite}$.

Doing standard perturbative calculations 
to first order in LV
coefficients and to first order in $\alpha$, 
the one-loop corrected wave-function renormalization and the fermion propagator pole%
, respectively 
read:
\begin{eqnarray}
\mathcal{Z}_R^{-1}  \label{Z_R} &= & \,\, 
(\mathcal{Z}_{R})_{\rm LI}^{-1}\,\, - \,\,  
\frac{2\alpha}{3\pi m^2}\left[
2c^p_p+ \tilde k^p_p\right],\\
\bar P(p) & = &\slash\!\!\! p + (c_{\rm phys})^p_\gamma - m_{\rm phys} +
\frac{\alpha}{3\pi m}\left(2(c_{\rm phys})^p_p-(\tilde k_{\rm phys})^p_p\right),
\label{P-phys}
\end{eqnarray}
where: 
%
%
\begin{equation}
m_{\rm phys} = m + \frac{\alpha m}{\pi} \left[  1 - \frac34 \gamma_E
-\frac34\ln\left(\frac{m^2}{4\pi\mu^2}\right) \right]
\label{m-phys}
\end{equation}
is the usual loop-corrected mass in the minimal-subtraction scheme,
and
\begin{equation}
(c_{\rm phys})^{\mu\nu}=c^{\mu\nu}
-\frac{\alpha}{3\pi}\left[\frac{29}{12}- \gamma_E - 
\ln\left(\frac{m^2}{4\pi\mu^2}\right)\right]
\bigl(2c^{\mu\nu}-\tilde k^{\mu\nu}\bigr).
\label{c-phys}
\end{equation}
Equation  \refeq{c-phys} expresses the radiatively corrected,
physical value of the Lorentz-violating parameter $c^{\mu\nu}$
in terms of its tree-level value.
Note that both $c^{\mu\nu}$ and the mass scale 
$\mu$  are unphysical,
renormalization-scheme-dependent quantities, unlike $(c_{\rm phys})^{\mu\nu}$, which is in principle measurable. This
is in fact granted by the cancellation between the terms containing $\mu$ in Eq. \refeq{c-phys} with
those coming from the running of $c^{\mu \nu}$\cite{KLP}. 
UV divergent quantities have been regularized with dimensional
regularization. 
Equations  \refeq{P-phys} and  \refeq{c-phys} show that LV radiative corrections 
depend only on $(2c^p_p-\tilde k^p_p)$ and $c^{\mu \nu}_{\rm phys}$ is indeed
infrared finite. Furthermore, it can be shown that the 
$\mathcal{O}(\alpha)$ LV radiative corrections to the 
dispersion relation 
are   proportional to 
$(2c^p_p-\tilde k^p_p)$ too. 
This property is a requirement coming from considerations of
field redefinitions which imply that, in this context, physically 
observable radiative effects should depend on $2c^{\mu \nu} - \tilde k^{\mu \nu}$ 
only\cite{fieldredef}.
Turning to the description of asymptotic states, 
Eq.\ \refeq{P-phys} implies a modified equation of motion 
for in- and out-spinors. Correspondingly 
an 
 adapted LSZ reduction formula for 
the fundamental scattering
 amplitude $\langle f | i \rangle$, controlled by
 Eqs.\ \refeq{Z_R} and \refeq{P-phys}, is obtained.

Our results allow us to conclude  that asymptotic single-particle states of fermions in 
Lorentz-violating QFT  receive  concrete
modifications due to radiative corrections. 
Specifically, the corresponding  Dirac equation turns out 
to be modified by Lorentz-violating operators not present 
in the Lagrangian, a novel feature of LV QFT as it does not occur in the Lorentz-symmetric case. Also, the fermion 
 wave-function renormalization is no longer a constant 
 but rather depends on the  LV coefficients under 
 consideration and on the external momentum as well. 
  As shown in Ref.\ \refcite{CLP} a non-trivial cancellation of IR divergences is also achieved in the fermion's dispersion relation and wave-function renormalization function. 
Furthermore,  IR divergences cancel in the  elastic 
Coulomb scattering cross section when the contribution of soft-photon emission from the corresponding external legs is taken into account.

\section*{Acknowledgments}
This work has been supported in part 
by: the Portuguese Funda\c c\~ao para a Ci\^encia e a Tecnologia,  the Mexican RedFAE, UNAB Grant DI-27-11/R, FONDECYT No. 11121633
and by the IUCSS.

\end{document}